\documentclass[aps,prd,reprint,preprintnumbers,groupedaddress]{revtex4-1}

\usepackage{natbib}
\usepackage{amsmath,amssymb}
\usepackage{graphicx}
\usepackage[utf8]{inputenc}
\usepackage{slashed}
\usepackage{bbm}

\begin{document}

\preprint{IPPP/16/103}

\title{How gauge covariance of the fermion and boson propagators in QED\\ constrain the effective fermion-boson vertex}

\author{Shaoyang Jia }
\email{sjia@email.wm.edu}
\affiliation{Physics Department, College of William \& Mary, Williamsburg, VA 23187, USA}
\author{ M.R. Pennington }
\email{michaelp@jlab.org}
\affiliation{Physics Department, College of William \& Mary, Williamsburg, VA 23187, USA ~and \\Theory Center, Thomas Jefferson National Accelerator Facility, Newport News, VA 23606, USA}



\date{\today}

\begin{abstract}
We derive the gauge covariance requirement imposed on the QED fermion-photon three-point function within the framework of a spectral representation for fermion propagators. When satisfied, such requirement ensures solutions to the fermion propagator Schwinger-Dyson equation (SDE) in any covariant gauge with arbitrary numbers of spacetime dimensions to be consistent with the Landau-Khalatnikov-Fradkin transformation (LKFT). The general result has been verified by the special cases of three and four dimensions. Additionally, we present the condition that ensures the vacuum polarization is independent of the gauge parameter. As an illustration, we show how the Gauge Technique dimensionally regularized in 4D does not satisfy the covariance requirement.
\end{abstract}


\maketitle

\section{Introduction}
The infinite set of Schwinger-Dyson equations constitutes the field equations of any theory. They relate Green's functions to each other. In QED and QCD, they relate, for instance, the fermion propagator to the gauge boson propagator and the fermion-boson interaction. The expansion of the Schwinger-Dyson equation for each Green's function in powers of the coupling yields the well-known perturbative series. However, most of the phenomena in hadron and nuclear physics are controlled by QCD in the regime of strong coupling \cite{Bashir:2012fs,Roberts:1994dr}. Then alternative truncations of the Schwinger-Dyson equations are required to capture the essence of the physics. A particularly simple truncation much used for the fermion propagator equation is to treat the fermion-boson vertex as simply proportional to $\gamma^\mu$ --- the Maris-Tandy ans\"atz \cite{Maris:1999nt}. It is argued that this might well be appropriate in the Landau gauge. Then the fermion mass function in the strong coupling regime can be shown to have a characteristic momentum dependence illustrated in Fig.~2 of Ref.~\cite{Williams:2006vva} (also see Fig.~1 of Ref.~\cite{Bhagwat:2003vw}). One realizes that whether the current quark mass (defined at some appropriate large momentum)  is 5 or 100 MeV, the mass at low momenta is $\sim 350$ MeV heavier: an infrared behavior that  matches constituent quark masses. It has been argued that such behavior of this gauge-dependent quantity is directly correlated with the momentum dependence of physical observables such as the electromagnetic formfactors of the pion and the proton \cite{Cloet:2013gva,Wilson:2011aa,Roberts:2015dea,Roberts:2015tha}. It would then seem natural to check how the mass functions shown in Ref.~\cite{Williams:2006vva} change with gauge. Solving the Schwinger-Dyson equation for the fermion in 4D with the same Maris-Tandy interaction in the Feynman gauge (for instance) changes the mass function as in Fig.~11 of Ref.~\cite{Kizilersu:2013hea}. The corresponding 3D results are illustrated by Fig.~3.2 of Ref.~\cite{Williams:2007zzh}.

\begin{figure}
\centering
\includegraphics[width=1\linewidth]{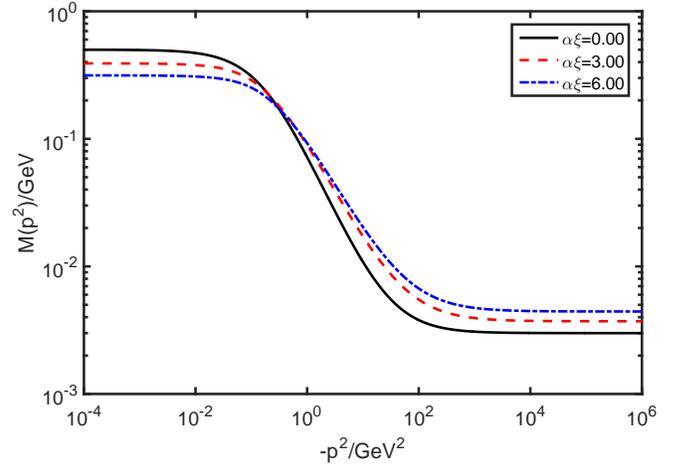}
\caption{The dependence of the fermion propagator mass function $M(p^2)$ on $\xi$. The black solid line is the parametric form of $M(p^2)$ given by Eq.~(2.1) of Ref.~\cite{Pennington:2009vz} with $M_0=3~\mathrm{MeV}$, $c=1.239$ and $\Lambda_{\mathrm{QCD}}=401~\mathrm{MeV}$. The red dashed line and the blue dash-dot line correspond to what the mass function should be when $\alpha\xi=3$ and $\alpha\xi=6$ respectively. The red dashed line and the blue dash-dot line are obtained by the LKFT for the fermion propagator in 4D within the $\overline{\mathrm{MS}}$ renormalization scheme at the scale $\mu=\Lambda$, with $F(p^2)=1$ and $M(p^2)$ given by the black line as the initial conditions in the Landau gauge.}
\label{fig:Mp2_xi_036}
\end{figure}
However, the gauge covariance of the fermion propagator is exactly specified by the Landau-Khalatnikov-Fradkin transformation (LKFT) \cite{Zumino:1959wt}. As we will remind the reader this relates the propagator functions in one gauge to those in another. If one applies this to the 4D fermion mass functions shown in Fig.~2 of Ref.~\cite{Williams:2006vva} for the Landau gauge, one obtains the behaviour in two other covariant gauges plotted in Fig.~\ref{fig:Mp2_xi_036}. The corresponding 3D results are given by Fig.~3.4 of Ref.~\cite{Williams:2007zzh}. The fact that the curves in Fig.~\ref{fig:Mp2_xi_036} strongly disagree with those in Fig.~11 of Ref.~\cite{Kizilersu:2013hea} indicates that the Maris-Tandy ans\"atz for the interaction cannot be appropriate in both the Landau and Feynman gauges. Indeed, it may not hold in any covariant gauge.

The purpose of this paper is to present the conditions that ensure the solutions of the Schwinger-Dyson equation (SDE) for the fermion propagator are gauge covariant \cite{Jia:2016udu}. We study this in QED, where particles having a physical mass-shell means it is natural that  the fermion propagator satisfies a spectral representation. This allows the SDE to be investigated at all momenta, and we are not restricted to spacelike momenta
(or nearby timelike momenta) as in QCD studies.

This article is organized as the following. In Section \ref{ss:spectral_rep}, the spectral representation is introduced for the fermion propagator. Then an abstract version of the SDE for the fermion propagator spectral functions is obtained in terms of the distribution $\Omega$. Section \ref{ss:LKFT} briefly reviews results of the LKFT for the fermion propagator. In Section \ref{ss:GC_requirements}, the gauge covariance requirements for the fermion propagator and the photon propagator SDEs are derived. 
In Section \ref{ss:decomposition_Omega}, with known contributions to $\Omega$ calculated, the consistency requirement for the fermion equation is written for the unknown terms of $\Omega$. As an example, the Gauge Technique anz\"{a}tz of Salam, Delbourgo and Strathdee~\cite{PhysRev.130.1287,PhysRev.135.B1398,PhysRev.135.B1428,Delbourgo:1977jc} translates into an $\Omega$ that is shown explicitly not to satisfy the consistency requirement in the quenched calculation in 4D. Section \ref{ss:summary} is the summary.
\section{Spectral representation of fermion propagator and its SDE\label{ss:spectral_rep}}
\subsection{Spectral representation of fermion propagator as a bijective mapping}
The fermion propagator $S_F(p)$ can be decomposed into Dirac vector and Dirac scalar components defined by
\begin{equation}
S_F(p)=S_1(p^2)\slashed{p}+S_2(p^2)\mathbbm{1}.
\end{equation}
Each component function is similar to a scalar propagator function. Therefore, the spectral representation of fermion propagator requires two scalar spectral functions;
\begin{equation}
S_{j}(p^2;\xi)=\int_{m^2}^{+\infty}ds\dfrac{\rho_{j}(s;\xi)}{p^2-s+i\varepsilon},\label{eq:fermion_spectral_rep}
\end{equation}
where $j=1,~2$ and the dependence on the covariant gauge parameter $\xi$ has been made explicit. The Feynman prescription of a momentum space propagator is denoted by the $i\varepsilon$ term in the denominator, while the normal writing of $\epsilon$ is reserved for how far away the number of spacetime dimensions is from 4 by $d=4-2\epsilon$. The spectral integral given in Eq.~\eqref{eq:fermion_spectral_rep} is convergent if $S_F(p)$ vanishes when $p^2\rightarrow \infty$.  That the integrals converge without the need for subtractions is assured
by the renormalizability of QED in $d < 4$ dimensions. 

Apparently when the fermion propagator takes its free-particle form, the spectral functions are given by $\rho_1(s)=\delta(s-m^2)$ and $\rho_2(s)=m\delta(s-m^2)$. When interactions are present, the fermion propagator is modified by quantum loop corrections and therefore develops branch cuts starting at the particle production thresholds. Such corrections add $\theta$-functions to the spectral functions. 
\begin{figure}
\includegraphics[width=0.7\linewidth]{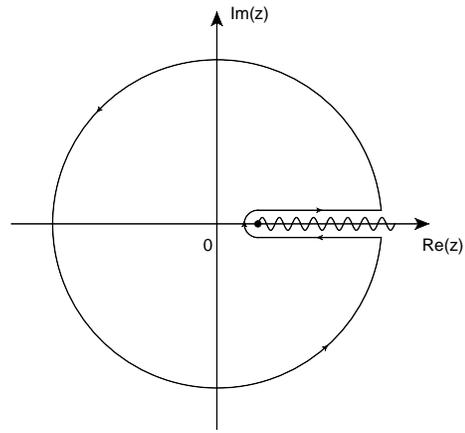}%
\caption{The illustration of analytic functions with branch cuts along the positive real axis. The contour can be used prove Eq.~\eqref{eq:fermion_spectral_rep} using Cauchy's integral formula. When used to prove Eq.~\eqref{eq:fermion_inv_spectral_rep}, $z$ stands for the complex $p^2$.}
\label{fig:analytic_fz}
\end{figure}

The existence of a spectral representation, Eq.~\eqref{eq:fermion_spectral_rep}, is determined by the analytic structure of the propagator functions $S_{j}(p^2)$ in the complex momentum plane. We expect that apart from free-particle poles and branch cuts along the positive real axis, propagator functions are holomorphic everywhere else. In this scenario, the spectral functions uniquely determine the propagator functions in the complex momentum plane. 

Meanwhile, when the analytic structure of the fermion propagator meets such requirements, the inverse of Eq.~\eqref{eq:fermion_spectral_rep} is given by 
\begin{equation}
\rho_{j}(s;\xi)=-\dfrac{1}{\pi}\mathrm{Im}\big\{S_{j}(s+i\varepsilon;\xi)\big\}.\label{eq:fermion_inv_spectral_rep}
\end{equation}
The Feynman prescription combined with the limiting form of the $\delta$-function, 
\[\lim\limits_{\varepsilon\rightarrow 0}\varepsilon/(x^2+\varepsilon^2)=\pi\delta(x),\]
specifies that any simple pole structure of the propagator function corresponds to a $\delta$-function term in its spectral function. In addition since $S_{j}(p^2+i\varepsilon)=S_{j}^*(p^2-i\varepsilon)$, functions calculated by Eq.~\eqref{eq:fermion_inv_spectral_rep} are indeed the spectral functions occurring in Eq.~\eqref{eq:fermion_spectral_rep}, which can be verified using Cauchy's integral formula with the contour in Fig.~\ref{fig:analytic_fz}. Therefore we have shown that the spectral representation given by Eq.~\eqref{eq:fermion_spectral_rep} and its inverse Eq.~\eqref{eq:fermion_inv_spectral_rep} specify a bijective mapping between the propagators as functions in the complex momentum plane and their spectral functions.
\subsection{SDE for fermion propagator spectral functions}
\begin{figure}
\includegraphics[width=\linewidth]{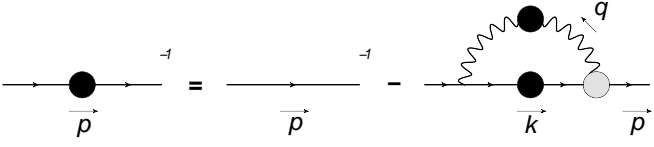}
\caption{The diagrammatic representation of the SDE for the fermion propagator in momentum space. The fermion-photon vertex is unknown and an ans\"atz is required to solve this equation.}
\label{fig:DSE_fermion_ori}
\end{figure}
The SDE for the fermion propagator in momentum space is represented by Fig.~\ref{fig:DSE_fermion_ori}. 
It has been solved extensively using specific ans\"{a}tze for the fermion photon vertex \cite{Curtis:1993py,Kizilersu:2014ela} (also see Ref.~\cite{Fukuda:1976zb,Atkinson:1993mz,Curtis:1992jg,Kizilersu:2000qd,Kizilersu:2001pd,Bloch:1995dd,Bashir:1994az,Bashir:2011dp,Kondo:1990ky,Kondo:1990st,Bashir:2011ij,Kizilersu:2013hea}). Solutions in the Minkowski space have been obtained by \cite{Maris:1994ux} (also see Ref.~\cite{Maris:1991cb,Atkinson:1989fp,Kusaka:1997xd,Sauli:2001we,Gutierrez:2016ixt}). Alternatively, these equations can be solved using complex conjugate poles to represent the propagator functions \cite{Bhagwat:2002tx,Bhagwat:2003vw,Raya:2015gva,Chang:2013pq}.

\begin{figure}
\includegraphics[width=\linewidth]{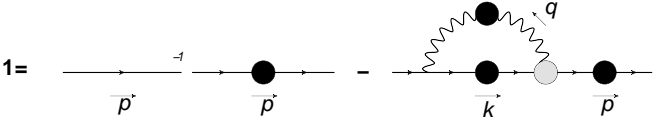}
\caption{The diagrammatic representation of the SDE for the fermion propagator spectral functions.}
\label{fig:DSE_fermion_rho}
\end{figure}

Each diagram in Fig.~\ref{fig:DSE_fermion_ori} is not obviously linear in the spectral functions $\rho_{j}(s;\xi)$. However, an easier way to solve for the spectral functions $\rho_{j}(s;\xi)$ directly from the propagator SDE is by multiplying each term of the equation depicted in Fig.~\ref{fig:DSE_fermion_ori} by $S_F(p;\xi)$. After this multiplication, we obtain Fig.~\ref{fig:DSE_fermion_rho}, where the first term on the right-hand side is clearly linear in $\rho_{j}(s;\xi)$. After decomposing this equation into its two Dirac components, the identity in Fig.~\ref{fig:DSE_fermion_rho} becomes
\begin{subequations}
\label{eq:SDE_fermion_linear_rho}
\begin{eqnarray}
& p^2S_1(p^2)-mS_2(p^2)+\sigma_1(p^2)=1\\
& S_2(p^2)-mS_1+\sigma_2(p^2)=0,
\end{eqnarray}
\end{subequations}
where $\sigma_{j}(p^2)$ are the Dirac scalar and vector components of the loop integral.
For the second term on the right-hand side of Fig.~\ref{fig:DSE_fermion_rho}, recall the Ward identity states that for QED, $Z_1=Z_2$ \cite{PhysRev.78.182}. Therefore the fermion propagator $S_F(p;\xi)$ shares the same renormalization constant with the fermion-photon vertex structure defined as $S_F(k)\Gamma^\mu(k,p)S_F(p)$, which indicates that the latter is also linear in $\rho_{j}(s;\xi)$. 

\begin{figure}
\centering
\includegraphics[width=1\linewidth]{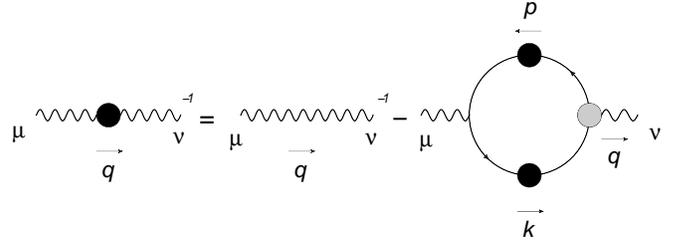}
\caption{The diagrammatic representation of the SDE for the photon propagator.}
\label{fig:SDE_photon}
\end{figure}
The gauge covariance of the solutions to the fermion and boson propagator Schwinger-Dyson equations will constrain the allowed forms of the fermion-boson vertex $\Gamma^\mu(k,p)$.
However the vertex in its full complexity with its 11 non-zero components is not required. Only the projections implied by the Schwinger-Dyson equation of
Figs.~\ref{fig:DSE_fermion_ori},~\ref{fig:DSE_fermion_rho}, and the corresponding equation for the inverse photon propagator ({\it i.e.} for the vacuum polarization) Fig.~\ref{fig:SDE_photon} are constrained. Thus it is this
{\it  effective} vertex that is restricted.  

One specific spectral construction of the vertex structure linear in $\rho_{j}(s;\xi)$ and satisfying the longitudinal Ward-Green-Takahashi identity is the Gauge Technique \cite{Delbourgo:1977jc}, which makes the ans\"{a}tz
\begin{equation}
S_F(p)\Gamma^\mu(k,p)S_F(p)=\int dW\dfrac{1}{\slashed{k}-W}\gamma^\mu\dfrac{1}{\slashed{p}-W}\,\rho(W),
\end{equation}
where $\rho(W)=\mathrm{sign}(W)[\rho_1(W^2)+W\rho_1(W^2)]$. In this particular case
\begin{align}
\quad\sigma_1(p^2)+\slashed{p}\sigma_2(p^2)&= ie^2\int d\underline{k}\int dW\,\rho(W)\nonumber\\
&\quad \times\gamma^\nu\dfrac{1}{\slashed{k}-W}\gamma^\mu\dfrac{1}{\slashed{p}-W}\,D_{\mu\nu}(q).\label{eq:sigma_12_oGT}
\end{align}
Transverse supplements to the Gauge Technique are required to meet various principles of QED, including renormalizablility \cite{Curtis:1990zs,Kizilersu:2009kg}, gauge covariance \cite{Bashir:2004mu} and transverse Ward-Green-Takahashi identities \cite{He:1999hb,He:2000we,He:2001cu,He:2006my}. However, from the equality $Z_1=Z_2$ \cite{PhysRev.78.182} we can further assume that such modifications are also linear in $\rho_{j}(s;\xi)$, and once known, allow us to calculate the loop integral in Fig.~\ref{fig:DSE_fermion_rho}, resulting in a function of $p$ as a linear functional of $\rho_j(s;\xi)$. Since this one-loop integral reduces to corrections to the fermion propagator in perturbative calculations, such $p^2$ dependences must be linearly generated from the free-particle propagator. Therefore after taking the imaginary part of Fig.~\ref{fig:DSE_fermion_rho}, or equivalently that of Eq.~\eqref{eq:SDE_fermion_linear_rho}, we obtain
\begin{subequations}
\label{eq:SDE_fermion_rho_s}
\begin{align}
s\rho_1(s;\xi)-m_B\rho_2(s;\xi)-\dfrac{1}{\pi}\mathrm{Im}\big\{ \sigma_1(s+i\varepsilon;\xi)\big\}=0,\label{eq:SDE_fermion_rho_1_s}\\
\rho_2(s;\xi)-m_B\rho_1(s;\xi)-\dfrac{1}{\pi}\mathrm{Im}\big\{\sigma_2(s+i\varepsilon;\xi)\big\}=0.\label{eq:SDE_fermion_rho_2_s}
\end{align}
\end{subequations}
The real constant term on the left-hand side disappears. After dividing Eq.~\eqref{eq:SDE_fermion_rho_1_s} by $s$, Eq.~\eqref{eq:SDE_fermion_rho_s} can be rewritten as
\begin{align}
\int ds'
\begin{pmatrix}
\Omega_{11}(s,s';\xi) & \Omega_{12}(s,s';\xi) \\ 
\Omega_{21}(s,s';\xi) & \Omega_{22}(s,s';\xi)
\end{pmatrix} 
\begin{pmatrix}
\rho_1(s';\xi) \\ 
\rho_2(s';\xi)
\end{pmatrix}&\nonumber\\
+
\begin{pmatrix}
\rho_1(s;\xi) \\ 
\rho_2(s;\xi)
\end{pmatrix}
=&
\begin{pmatrix}
0 \\ 
0
\end{pmatrix},
\label{eq:SDE_fermion_rho_itg}
\end{align}
where the $\Omega_{ij}(s,s';\xi)$ encode all required linear operations on the spectral functions $\rho_{j}(s;\xi)$, which are obtained by functional derivatives similar to
\begin{equation}
\Omega(s,s')=-\dfrac{\delta}{\delta \rho(s')}\dfrac{1}{\pi}\mathrm{Im}\big\{\sigma(s+i\varepsilon)\big\}.\label{eq:delta_delta_rho}
\end{equation}
The bare mass coupling in Eq.~\eqref{eq:SDE_fermion_rho_s} is explicitly included in the off-diagonal terms of $\Omega_{ij}(s,s';\xi)$. When the fermion-photon vertex is given by the Gauge Technique the resulting $\sigma_{j}$ is given by Eq.~\eqref{eq:sigma_12_oGT}. Then $m_B$ is the only coupling between equations for $\rho_1$ and $\rho_2$. However, when dimension-odd operators are allowed to enter the expression for $S_F(k)\Gamma^\mu(k,p)S_F(p)$, they will contribute additionally to off-diagonal elements of $\Omega_{ij}$.

For a given ans\"atz for the fermion-photon vertex that ensures $S_F(k)\Gamma^\mu(k,p)S_F(p)$ being linear in $\rho_{j}(s;\xi)$, there is a corresponding $\Omega$. It is the matrix $\Omega$ that is constrained by gauge covariance. Regardless of the photon being quenched or not, the SDE for fermion propagator spectral functions takes the form of Eq.~\eqref{eq:SDE_fermion_rho_itg}. Solutions to Eq.~\eqref{eq:SDE_fermion_rho_itg} found in different covariant gauges are, of course, different because the fermion propagator is not a physical observable. However any ans\"{a}tz for the fermion-photon vertex that respects Eq.~\eqref{eq:SDE_fermion_rho_itg} is expected to be gauge covariant. Satisfying the Ward-Green-Takahashi identity, a consequence of gauge invariance, however, is not sufficient to ensure the gauge covariance of solutions to Eq.~\eqref{eq:SDE_fermion_rho_itg}, as we will see explicitly later on. 
In order to explore the conditions on the  $\Omega_{ij}(s,s';\xi)$ that ensure gauge covariance of solutions to Eq.~\eqref{eq:SDE_fermion_rho_itg}, the LKFT for the fermion propagator needs to be solved first. 
\section{LKFT for fermion propagator\label{ss:LKFT}}
Detailed discussion of the LKFT for the fermion propagator has been made elsewhere \cite{Jia:2016wyu}. Consequently in the present article, only crucial intermediate steps are included. Because of the existence of bijective relations among the fermion propagator in coordinate space, in momentum space and in spectral representation, LKFT manifests itself as isomorphic representations for the fermion propagator in these spaces. Being linear in the coordinate representation suggests that LKFT in the spectral representation should also be a linear transform. Therefore without loss of generality,
\begin{equation}
\rho_{j}(s;\xi)=\int ds' \mathcal{K}_{j}(s,s';\xi)\rho_{j}(s';0),\label{eq:LKFT_linearity_spectral_rep}
\end{equation}
where distributions $\mathcal{K}_{j}(s,s';\xi)$ represent linear operations that encode $\xi$ dependences of $\rho_{j}(s;\xi)$ to be determined by LKFT. These operations observe closure, associativity, the existence of the identity element and the inverse elements.

Because the LKFT is independent of the initial conditions for $\rho_{j}(s;\xi)$, one can obtain the following differential equations for $\mathcal{K}_{j}(s,s';\xi)$ by taking the $\xi$ derivative of the coordinate space LKFT and subsequently taking the Fourier transform, 
\begin{equation}
\dfrac{\partial}{\partial\xi}\int ds\,\dfrac{\mathcal{K}_{j}(s,s';\xi)}{p^2-s+i\varepsilon}=-\dfrac{\alpha}{4\pi}\int ds\,\dfrac{\Xi_{j}(p^2,s)}{p^2-s+i\varepsilon}\,\mathcal{K}_{j}(s,s';\xi),\label{eq:LKFT_k12}
\end{equation}
where $z=p^2/s$. Explicit calculations reveal that
\begin{subequations}\label{eq:Xi_reduced}
\begin{align}
\dfrac{\Xi_1}{p^2-s}&=\dfrac{\Gamma(\epsilon)}{s}\left(\dfrac{4\pi\mu^2}{s}\right)^\epsilon\dfrac{(-2)~_2F_1(\epsilon+1,3;3-\epsilon;z)}{(1-\epsilon)(2-\epsilon)}\label{eq:Xi_1_reduced}\\
\dfrac{\Xi_2}{p^2-s}&=\dfrac{\Gamma(\epsilon)}{s}\left(\dfrac{4\pi\mu^2}{s}\right)^\epsilon\dfrac{-1}{1-\epsilon}~_2F_1(\epsilon+1,2;2-\epsilon;z).\label{eq:Xi_2_reduced}
\end{align}
\end{subequations}
where recall $\epsilon$ is defined by $d = 4 - 2 \epsilon$.
Eq.~\eqref{eq:LKFT_k12} is solved by 
\begin{equation}
\mathcal{K}_{j}=\exp\left(-\dfrac{\alpha\xi}{4\pi}\Phi_{j}\right),\label{eq:k_exponential}
\end{equation}
where $\Phi_{j}$ are distributions independent of $\xi$ and can be solved from
\begin{equation}
\int ds\dfrac{\Phi_{j}(s,s')}{p^2-s+i\varepsilon}=\dfrac{\Xi_{j}(p^2,s')}{p^2-s'+i\varepsilon}.\label{eq:Phi_Xi}
\end{equation}
Eq.~\eqref{eq:Phi_Xi} is solved once we have established how to generate the $z$ dependences in Eq.~\eqref{eq:Xi_reduced} from linear operations on the free-particle propagator with respect to the variable $s$ alone. To do so requires the Riemann-Liouville definition of fractional calculus \cite{Riemann:fractional};
\begin{equation}
	I^\alpha f(z)=\dfrac{1}{\Gamma(\alpha)}\int_{0}^{z}dz'(z-z')^{\alpha-1}f(z').\label{eq:def_Riemann_Liouville_I}
\end{equation}
For $\alpha>0$, the Riemann-Liouville fractional derivative is defined as
\begin{equation}
	D^\alpha f(z)=\left(\dfrac{d}{dz}\right)^{\lceil \alpha \rceil}I^{\lceil\alpha\rceil-\alpha}f(z),
\end{equation}
where $\lceil \alpha \rceil$ is the ceiling function. Specifically for ${\alpha \in (0,1)}$, $\lceil \alpha \rceil=1$ and
\begin{equation}
	D^\alpha f(z)=\dfrac{1}{\Gamma(1-\alpha)}\dfrac{d}{dz}\int_{0}^{z}dz'(z-z')^{-\alpha}f(z').\label{eq:def_Riemann_Liouville_D}
\end{equation}
Furthermore, we define the dimensionless operator $\phi$ such that at the operator level $\int ds'\Phi=\phi$, then
\begin{equation}
\phi_n=\Gamma(\epsilon)\left(\dfrac{4\pi\mu^2}{p^2}\right)^\epsilon \dfrac{\Gamma(1-\epsilon)}{\Gamma(1+\epsilon)}z^{2\epsilon+2-n}D^\epsilon z^{n-1}D^\epsilon z^{\epsilon-1}.\label{eq:def_phi}
\end{equation}
Distributions $\phi_n$ in Eq.~\eqref{eq:def_phi} correspond to $\Phi_{j}$ with ${n=3,~2}$ for ${j=1,~2}$; see Eqs.~(\ref{eq:Xi_1_reduced},~\ref{eq:Xi_2_reduced}) respectively.

The exponential form of $\mathcal{K}_{j}$ given by Eq.~\eqref{eq:k_exponential} remains illusive even with $\phi_n$ explicitly written as Eq.~\eqref{eq:def_phi}. To see how $\mathcal{K}_{j}$ works explicitly, consider any function of the spectral variable that can be written as a linear combination of $z^\beta$, we can show that
	\begin{align}
	&\quad \mathcal{K}_{j}z^\beta 
	=\exp\left(-\overline{\alpha}\overline{\phi}_n \right)z^\beta=\sum_{m=0}^{+\infty}\dfrac{(-\overline{\alpha})^m}{m!}\overline{\phi}^m_nz^\beta\nonumber\\
	& =\sum_{m=0}^{+\infty}\dfrac{(-\overline{\alpha})^m}{m!}\dfrac{\Gamma(n+\beta+(m-1)\epsilon-1)\Gamma(\beta+m\epsilon)}{\Gamma(n+\beta-\epsilon-1)\Gamma(\beta)}z^{\beta+m\epsilon},\label{eq:kn_zbeta}
	\end{align}
where 
	\begin{equation}
	\overline{\alpha}\equiv \dfrac{\alpha\xi}{4\pi}\dfrac{\Gamma(\epsilon)\Gamma(1-\epsilon)}{\Gamma(1+\epsilon)}\left(\dfrac{4\pi\mu^2}{p^2}\right)^\epsilon.\label{eq:def_alpha}
	\end{equation}
Specifically for small $\epsilon$, the operations given by Eq.~\eqref{eq:kn_zbeta} reduce to
\begin{align}
\mathcal{K}_j& =\left(\dfrac{\mu^2z}{p^2} \right)^{-\nu}\exp\bigg\{-\nu\left[\dfrac{1}{\epsilon}+\gamma_E+\ln 4\pi+\mathcal{O}(\epsilon^1) \right] \bigg\}\nonumber\\
& \quad \times z^{2-n}I^{\nu}z^{n-1-\nu}I^{\nu}z^{-\nu-1},\label{eq:kn_small_epsilon}
\end{align}
where $\nu=\alpha\xi/(4\pi)$.
\section{Gauge covariance requirements for the propagator SDEs\label{ss:GC_requirements}}
\subsection{Gauge covariance requirement on the fermion propagator SDE}
For notational convenience, when two distributions are multiplied together, the integration over the spectral variable is implied. After adopting this notation, only dependences on $\xi$ are required to be written explicitly. Therefore Eq.~\eqref{eq:SDE_fermion_rho_itg} becomes
\begin{equation}
\begin{pmatrix}
\rho_1(\xi) \\ 
\rho_2(\xi)
\end{pmatrix} +
\begin{pmatrix}
\Omega_{11}(\xi) & \Omega_{12}(\xi) \\ 
\Omega_{21}(\xi) & \Omega_{22}(\xi)
\end{pmatrix} 
\begin{pmatrix}
\rho_1(\xi) \\ 
\rho_2(\xi)
\end{pmatrix}=
\begin{pmatrix}
0 \\
0
\end{pmatrix}
.\label{eq:DSE_xi_matrix}
\end{equation}
Since LKFT does not couple $\rho_1$ with $\rho_2$, we have the following abbreviated versions of Eq.~\eqref{eq:LKFT_linearity_spectral_rep},
\begin{equation}
\rho_j(\xi)=\mathcal{K}_j(\xi)\rho_j(0).\label{eq:LKFT_linearity_spectral_rep_abb}
\end{equation}
Substituting Eq.~\eqref{eq:LKFT_linearity_spectral_rep_abb} into Eq.~\eqref{eq:DSE_xi_matrix} gives
\begin{align}
&\begin{pmatrix}
\mathcal{K}_{1}(\xi) & \\ 
 & \mathcal{K}_{2}(\xi)
\end{pmatrix} 
\begin{pmatrix}
\rho_1(0) \\ 
\rho_2(0)
\end{pmatrix}+\nonumber\\
&
\begin{pmatrix}
\Omega_{11}(\xi) & \Omega_{12}(\xi) \\ 
\Omega_{21}(\xi) & \Omega_{22}(\xi)
\end{pmatrix} 
\begin{pmatrix}
\mathcal{K}_1(\xi) &  \\ 
 & \mathcal{K}_2(\xi)
\end{pmatrix} 
\begin{pmatrix}
\rho_1(0) \\ 
\rho_2(0)
\end{pmatrix}=
\begin{pmatrix}
0\\ 
0
\end{pmatrix}
.\label{eq:DSE_xi_matrix_rho0}
\end{align}
Since obviously \[(\mathrm{diag}\{\mathcal{K}_1(\xi),~\mathcal{K}_2(\xi) \})^{-1}=\mathrm{diag}\{\mathcal{K}_1(-\xi),~\mathcal{K}_2(-\xi) \}\] with matrix inversion defined by regular matrix multiplication and distribution inversion defined by distribution multiplication that gives a $\delta$-function. Combining this result with Eq.~\eqref{eq:DSE_xi_matrix} in the Landau gauge,
\begin{equation}
\begin{pmatrix}
\rho_1(0) \\ 
\rho_2(0)
\end{pmatrix} +
\begin{pmatrix}
\Omega_{11}(0) & \Omega_{12}(0) \\ 
\Omega_{21}(0) & \Omega_{22}(0)
\end{pmatrix} 
\begin{pmatrix}
\rho_1(0) \\ 
\rho_2(0)
\end{pmatrix}=
\begin{pmatrix}
0 \\
0
\end{pmatrix}
,\label{eq:DSE_0_matrix}
\end{equation}
yields
\begin{align}
&\begin{pmatrix}
\Omega_{11}(0) & \Omega_{12}(0) \\ 
\Omega_{21}(0) & \Omega_{22}(0)
\end{pmatrix} =\nonumber\\
& 
\begin{pmatrix}
\mathcal{K}_1(-\xi) & \\ 
 & \mathcal{K}_2(-\xi)
\end{pmatrix} 
\begin{pmatrix}
\Omega_{11}(\xi) & \Omega_{12}(\xi) \\ 
\Omega_{21}(\xi) & \Omega_{22}(\xi)
\end{pmatrix} 
\begin{pmatrix}
\mathcal{K}_1(\xi) & \\ 
 & \mathcal{K}_2(\xi)
\end{pmatrix}.\label{eq:consistency_fermion_prop_SDE_LKFT}
\end{align}
Since for different ans\"atz the Landau gauge solutions $\rho(s;0)$ are allowed to be different, Eq.~\eqref{eq:consistency_fermion_prop_SDE_LKFT} is the necessary condition for solutions to the SDE for the fermion propagator to be consistent with its LKFT. 

Meanwhile, when $\Omega(0)$ is given by Eq.~\eqref{eq:consistency_fermion_prop_SDE_LKFT}, Eq.~\eqref{eq:DSE_0_matrix} becomes Eq.~\eqref{eq:DSE_xi_matrix_rho0}, which, when viewed as equations for $\mathcal{K}_1(\xi)\rho_1(0)$ and $\mathcal{K}_2(\xi)\rho_2(0)$, is identical to Eq.~\eqref{eq:DSE_xi_matrix}. Therefore Eq.~\eqref{eq:consistency_fermion_prop_SDE_LKFT} is also the sufficient condition for solutions to the fermion propagator SDE to be consistent with LKFT. Therefore solutions of the SDE for fermion propagator are consistent with LKFT if and only if Eq.~\eqref{eq:consistency_fermion_prop_SDE_LKFT} is satisfied.
\subsection{Gauge covariance requirement on the photon propagator SDE\label{ss:GC_requirement_photon}}
After gauge fixing, the photon propagator becomes
\begin{equation}
D^{\mu\nu}(q)=\Delta^{\mu\nu}(q)+\xi\dfrac{q^\mu q^\nu}{q^4+i\varepsilon},\label{eq:photon_prop}
\end{equation}
where
\begin{equation}
\Delta^{\mu\nu}(q)=\dfrac{G(q^2)}{q^2+i\varepsilon}\left(g^{\mu\nu}-\dfrac{q^\mu q^\nu}{q^2}\right)\label{eq:transverse_photon}
\end{equation}
is the Landau gauge photon propagator. The dressing function $G(q^2)$ is determined by the SDE for the photon propagator.

As illustrated in Fig.~\ref{fig:SDE_photon}, the same vertex structure $S_F(k)\Gamma^\mu(k,p)S_F(p)$ appears in the SDE for photon propagator. This allows us to derive the gauge covariance requirement on the photon propagator SDE. Meanwhile, the spectral representation ensures the transversality of the vacuum polarization tensor through the translational invariance of the loop momentum. 
To start with, the dependence of the photon propagator $D^{\mu\nu}(q)$ on the covariant gauge parameter $\xi$ is completely specified by the $\xi q^\mu q^\nu/q^4$ term, as a direct consequence of which, $G(q^2)$ of Eq.~\eqref{eq:transverse_photon} and the transverse vacuum polarization tensor $\Pi^{\mu\nu}(q^2)=(g^{\mu\nu}q^2-q^\mu q^\nu)\Pi(q^2)$ are required to be independent of $\xi$.

Next, one should expect the analytic structure of the photon propagator to differ from that for the fermion propagator, with singularities in distinct (but related) positions. Nevertheless, we can still proceed by keeping the external momentum dependence explicit without introducing a spectral function for the photons. Therefore the consistency requirement for the photon propagator SDE is simply given by $\partial_\xi \Pi(q^2)=0$. Since the vacuum polarization function $\Pi(q^2)$ is linear in the fermion propagator spectral functions. One can write
\begin{align}
&\quad \Pi(q^2) =\int dW\,\Omega^\gamma(q^2,W;\xi)\,\rho(W;\xi)\nonumber\\
& =\int ds\, (\Omega^\gamma_1(q^2,s;\xi),~\Omega^\gamma_2(q^2,s;\xi))\,
\begin{pmatrix}
\rho_1(s;\xi) \\ 
\rho_2(s;\xi)
\end{pmatrix}. 
\end{align}
With the $\xi$ dependence of $\rho_j(s;\xi)$ given by Eq.~\eqref{eq:k_exponential}, the $\xi$ independence of $\Pi(q^2)$ specifies
\begin{equation}
\Omega_{j}^\gamma(q^2,s;\xi)=\int ds'\,\Omega_{j}^\gamma(q^2,s';0)\,\exp\left[\dfrac{\alpha\xi}{4\pi}\Phi_{j}(s',s) \right],\label{eq:consistency_gamma}
\end{equation}
or at the operator level $\Omega_\xi^\gamma=\Omega_0^\gamma e^{\nu\Phi}$. This is the consistency requirement between the photon SDE and the LKFT.
\section{The decomposition of $\Omega$\label{ss:decomposition_Omega}}
The operator $\Omega$ can be decomposed into components from the fermion mass, and the longitudinal and transverse parts of the photon propagator. Some of these can be calculated exactly. In the quenched approximation, $G(q^2)=1$ and the photon propagator is known exactly. When photons are unquenched, the vacuum polarization produces a nontrivial $G(q^2)$ in Eq.~\eqref{eq:transverse_photon}. In this case the introduction of a spectral representation for the photon propagator is required. Meanwhile, since the longitudinal part of the fermion-photon vertex is fixed by the Ward-Green-Takahashi identity, contributions from the $\xi q^\mu q^\nu/q^4$ term to $\Omega$ are known exactly regardless of either the dressing of the photon propagator or the transverse part of the fermion-photon vertex. 

While the bare mass $m_B$ contributes to off-diagonal terms of $\Omega_{ij}$ containing terms at most linear in $m_B$, allowing the following decomposition of $\Omega$,
\begin{equation}
\Omega=\Omega^m+\Omega^\xi+\Omega^\Delta,\label{eq:Omega_decomposition}
\end{equation}
where
\begin{equation}
\Omega^m(s,s')=
\begin{pmatrix}
 & -\dfrac{m_B}{s}\delta(s-s') \\ 
-m_B\delta(s-s') & 
\end{pmatrix} 
\end{equation}
stands for the operation linear in $m_B$ that is also independent of $\xi$. Furthermore, denoting by $\Omega^\xi$ the contribution from the longitudinal component of the photon propagator $\xi q^\mu q^\nu/q^4$, this can be readily computed exactly. While $\Omega^\Delta$ is calculated with the $\Delta^{\mu\nu}(q)$ term of the photon propagator in Eq.~\eqref{eq:photon_prop}, which remains unknown without either the photon dressing function or the transverse part of the fermion-photon vertex. 

$\Omega^\xi$, being linear in $\xi$, vanishes in the Landau gauge. While $\Omega^\Delta$ depends on the gauge because of the transverse aspects of $S_F(k)\Gamma^\mu(k,p)S_F(p)$. These need not be zero in the Landau gauge, despite this being commonly assumed.
\subsection{Exact expressions for $\Omega^\xi$}
In order to calculate $\Omega^\xi$ in any dimensions, based on Eq.~\eqref{eq:delta_delta_rho} we need to calculate the contribution to $\sigma_j(p^2;\xi)$ as functionals of $\rho_j(s;\xi)$ with explicit dependence on the number of spacetime dimensions $d=4-2\epsilon$. We denote by $\sigma_j^\xi$ the contribution to $\sigma_j$ from the longitudinal component of the photon propagator. After replacing $D_{\mu\nu}(q)$ by $\xi q_\mu q_\nu/q^4$, we have
\begin{widetext}
\begin{align}
& \quad \sigma_1^{\xi}(p^2)+\slashed{p}\sigma_2^{\xi}(p^2)=ie^2\xi\int dW\int d\underline{k}\,\slashed{q}\,\dfrac{1}{\slashed{k}-W}\,\slashed{q}\,\dfrac{1}{q^4}\,\dfrac{\rho(W)}{\slashed{p}-W}\nonumber\\
& =\dfrac{-\alpha\xi}{4\pi}\int dW\int_{0}^{1}dy\,\Gamma(\epsilon)\left(\dfrac{4\pi\mu^2}{s} \right)^\epsilon\Bigg\{\dfrac{-\epsilon(1-y)y}{(1-yz)\mathcal{D}^\epsilon}\slashed{p}+\dfrac{y[3\epsilon-4+(3-2\epsilon)y]}{\mathcal{D}^\epsilon}\slashed{p}+\dfrac{W}{\mathcal{D}^\epsilon}\Bigg\}\dfrac{\rho(W)}{\slashed{p}-W},\label{eq:def_sigma_overline_Delta}
\end{align}
with $y$ being the Feynman parameter, $z=p^2/s$ and the combined denominator given by $\mathcal{D}=(1-y)(1-y z)$. After applying the integral definition of hypergometric functions \cite{abramowitz1964handbook}, we have 
\begin{equation}
\int_{0}^{1}dy\,\dfrac{y^p(1-y)^q}{(1-yz)^a}=\dfrac{\Gamma(p+1)\Gamma(q+1)}{\Gamma(p+q+2)}~_2F_1(a,p+1;p+q+2;z).
\end{equation}
Then the loop-integral factor of Eq.~\eqref{eq:def_sigma_overline_Delta} becomes
\begin{align}
ie^2\int d\underline{k}\,\slashed{q}\,\dfrac{1}{\slashed{k}-W}\,\dfrac{\slashed{q}}{q^4}
& = -\dfrac{\alpha}{4\pi}\Gamma(\epsilon)\left(\dfrac{4\pi\mu^2}{s} \right)^\epsilon\Bigg\{\dfrac{-\epsilon\slashed{p}}{(3-\epsilon)(2-\epsilon)}~_2F_1(1+\epsilon,2;4-\epsilon;z)+\dfrac{(3\epsilon-4)\slashed{p}}{(2-\epsilon)(1-\epsilon)}~_2F_1(\epsilon,2;3-\epsilon;z)+\nonumber\\
& \quad \dfrac{2(3-2\epsilon)\slashed{p}}{(3-\epsilon)(2-\epsilon)(1-\epsilon)}~_2F_1(\epsilon,3;4-\epsilon;z)+\dfrac{W}{1-\epsilon}~_2F_1(\epsilon,1;2;z)  \Bigg\}.\label{eq:sigma_overline_delta_loop}
\end{align}
\end{widetext}
Since $\sigma_{j}^{\xi}$ are properly formulated Feynman diagrams corresponding to loop-corrections to the fermion propagator where the $\rho_j(s;\xi)$ are given by $\delta$-functions, one expects that linear combinations of hypergeometric functions in Eq.~\eqref{eq:sigma_overline_delta_loop} are finite (at least in 4D) when $z\rightarrow 1$ such that there are contributions to fermion propagator functions no more singular than those of a free particle.

After numerous applications of contiguous relations for hypergeometric functions $~_2F_1(a,b;c;z)$, Eq.~\eqref{eq:def_sigma_overline_Delta} becomes
\begin{subequations}\label{eq:sigma_xi}
\begin{align}
& \sigma_1^{\xi}(p^2)=\dfrac{\alpha\xi}{4\pi}\int ds\left(\dfrac{4\pi\mu^2}{s} \right)^\epsilon \dfrac{\Gamma(\epsilon)}{1-\epsilon}\nonumber\\
&\hspace{2cm} \times~_2F_1(\epsilon,2;2-\epsilon;z)\,\rho_1(s),\label{eq:sigma_1_xi}\\
& \sigma_2^{\xi}(p^2)=\dfrac{\alpha\xi}{4\pi}\int ds\left(\dfrac{4\pi\mu^2}{s} \right)^\epsilon \dfrac{\epsilon\Gamma(\epsilon)}{(2-\epsilon)(1-\epsilon)}\nonumber\\
& \hspace{2cm}\times~_2F_1(\epsilon+1,2;3-\epsilon;z)\dfrac{1}{s}\,\rho_2(s).\label{eq:sigma_2_xi}
\end{align}
\end{subequations}
Details of the intermediate steps can be found in Appendix \ref{ss:simplify_sigma_xi}.

Next, since $\Omega^\xi$ is only linear in $\xi$, we define $\Theta$ as 
\begin{equation}
\Theta=-\nu^{-1}\Omega^\xi \; ,\label{eq:def_Theta}
\end{equation}
(recalling $\nu\equiv \alpha\xi/4\pi$) such that the distribution $\Theta$ is independent of $\xi$. Apparently only diagonal elements of $\Theta_{ij}$ survive, therefore
\begin{subequations}\label{eq:def_Theta_dd}
\begin{align}
&  -\nu\Theta_{11}(s,s';\xi)=-\dfrac{1}{s}\dfrac{\delta}{\delta\rho_1(s';\xi)}\dfrac{1}{\pi}\mathrm{Im}\big\{\sigma_1^\xi(s+i\varepsilon;\xi)\big\}.\label{eq:def_Theta_11}\\
& -\nu\Theta_{22}(s,s';\xi)=-\dfrac{\delta}{\delta\rho_2(s';\xi)}\dfrac{1}{\pi}\mathrm{Im}\big\{\sigma_2^\xi(s+i\varepsilon;\xi)\big\}.\label{eq:def_Theta_22}
\end{align}
\end{subequations}
Let us define at the operator level $\int ds\Theta=\mathrm{diag}\big\{\theta_1,\theta_2\big\}$. Since
\[\dfrac{1}{p^2-s+i\varepsilon}=-\dfrac{1}{p^{2}}\sum_{\beta=1}^{+\infty}z^\beta,\] Eqs.~(\ref{eq:sigma_xi},~\ref{eq:def_Theta_dd}) imply
\begin{subequations}
\begin{align}
-\nu\theta_1\dfrac{1}{p^2-s+i\varepsilon}& =\nu\Gamma(\epsilon)\left(\dfrac{4\pi\mu^2}{p^2} \right)^\epsilon\dfrac{1}{p^2}\dfrac{1}{1-\epsilon}\nonumber\\
&\quad\times\sum_{\beta=1}^{+\infty}\dfrac{(\epsilon)_{\beta-1}(2)_{\beta-1}}{(2-\epsilon)_{\beta-1}(\beta-1)!}z^{\beta+\epsilon}\\
-\nu\theta_2\dfrac{1}{p^2-s+i\varepsilon}& =\nu\Gamma(\epsilon)\left(\dfrac{4\pi\mu^2}{p^2} \right)^\epsilon\dfrac{1}{p^2}\dfrac{\epsilon}{(2-\epsilon)(1-\epsilon)}\nonumber\\
& \quad\times \sum_{\beta=1}^{+\infty}\dfrac{(\epsilon+1)_{\beta-1}(2)_{\beta-1}}{(3-\epsilon)_{\beta-1}(\beta-1)!}z^{\beta+\epsilon}.
\end{align}
\end{subequations}
Therefore we have the following identities for $\theta_j$,
\begin{subequations}
\begin{align}
& \theta_1 z^\beta=\left(\dfrac{4\pi\mu^2}{p^2} \right)^\epsilon\dfrac{\Gamma(1-\epsilon)\Gamma(\beta+\epsilon-1)\beta}{\Gamma(1+\beta-\epsilon)}z^{\beta+\epsilon},\\
& \theta_2 z^\beta=\left(\dfrac{4\pi\mu^2}{p^2}\right)^\epsilon\dfrac{\Gamma(1-\epsilon)\Gamma(\beta+\epsilon)\beta}{\Gamma(2+\beta-\epsilon)}z^{\beta+\epsilon},
\end{align}
\end{subequations}
which completely specify $\Theta$, and consequently $\Omega^\xi$.
\subsection{Consistency requirement as recurrence relations}
Based on previous analysis, for a given ans\"atz for the fermion-photon vertex that ensures the vertex structure $S_F(k)\Gamma^\mu(k,p)S_F(p)$ being linear in $\rho(W)$, the corresponding distributions $\Omega_{ij}$ can be calculated. Such an ans\"atz is consistent with LKFT if and only if Eq.~\eqref{eq:consistency_fermion_prop_SDE_LKFT} is satisfied. Independent of any ans\"atz, two terms $\Omega^m$ and $\Omega^\xi$ are now known exactly. 

In this subsection we explore how Eq.~\eqref{eq:consistency_fermion_prop_SDE_LKFT} is satisfied incorporating $\Omega^\Delta$, \textit{i.e.} with $\Omega^m$ and $\Omega^{\xi}$ explicitly included. Straightforwardly, one could substitute Eq.~\eqref{eq:Omega_decomposition} with known components into the consistency requirement Eq.~\eqref{eq:consistency_fermion_prop_SDE_LKFT}, and obtain 
\begin{equation}
\Omega^\Delta=e^{-\nu\Phi}(\Omega^m+\Omega^\Delta_0)e^{\nu\Phi}-\Omega^m+\nu\Phi,
\end{equation}
as the consistency requirement on $\Omega^\Delta$. Alternatively, with LKFT for fermion propagator spectral functions given by Eq.~\eqref{eq:k_exponential}, we have
\begin{equation}
\Omega_\xi=e^{-\nu\Phi}\Omega_0 e^{\nu\Phi},\label{eq:Omega_xi_0}
\end{equation}
where the subscript of $\Omega_\xi$ highlights the $\xi$ dependent $\Omega$ in Eq.~\eqref{eq:Omega_decomposition}, therefore $\Omega_0=\lim\limits_{\xi\rightarrow 0}\Omega_\xi$. To see how infinitesimal changes in $\xi$ affect $\Omega^\Delta$, consider taking the derivative with respect to $\nu$ (effectively $\xi$) of Eq.~\eqref{eq:Omega_xi_0},
\[\partial_\nu\Omega_\xi=-\Phi e^{-\nu\Phi}\Omega_0 e^{\nu\Phi}+  e^{-\nu\Phi}\Omega_0 e^{\nu\Phi}\Phi=[\Omega_\xi,\Phi].\]
Substituting in Eq.~\eqref{eq:Omega_decomposition} and Eq.~\eqref{eq:def_Theta} produces
\begin{equation}
\partial_\nu\Omega^\Delta+[\Phi,\Omega^\Delta]=-[\Phi,\Omega^m]+\Theta+\nu[\Phi,\Theta].\label{eq:D_xi_Omega_Delta}
\end{equation}
In order to recover the corresponding terms using the spectral representation for the fermion propagator, one calculates
\begin{equation}
\int ds\,ds'\dfrac{1}{p^2-s+i\varepsilon}\,\Omega(s,s')\, \rho(s').\label{eq:convention_S_rho}
\end{equation}
Since the $z^\beta$ expansion is in fact the $p^2/s$ expansion of the free-particle propagator, commutators of operations on $z^\beta$ should be calculated with $z^\beta$ to the left. There exists an alternative convention to Eq.~\eqref{eq:convention_S_rho} that locates the free-particle propagator to the right of the operation, which subsequently modifies Eq.~\eqref{eq:consistency_fermion_prop_SDE_LKFT}. The net effect of adopting the alternative convention to solutions of Eq.~\eqref{eq:D_xi_Omega_Delta} is, however, zero compared with the convention given by Eq.~\eqref{eq:convention_S_rho} because deriving Eq.~\eqref{eq:consistency_fermion_prop_SDE_LKFT} using the alternative convention for the location of free-particle propagator leads to exchanging $\mathcal{K}_j^\xi$ with $\mathcal{K}_j^{-\xi}$.

Within this convention of locations, the right-hand side of Eq.~\eqref{eq:D_xi_Omega_Delta}, operating on $z^\beta$ can be calculated according to Eqs.~(\ref{eq:zbeta_phi_theta},~ \ref{eq:zbeta_phi}).

Since physical $\Omega^\Delta$ are generated by loop-corrections to the fermion propagator, the following criteria apply:
\begin{itemize}
\item while the dependence of $\Omega^\Delta$ on $\nu=\alpha\xi/(4\pi)$ is allowed to be any order, $\Omega^\Delta$ cannot depend on the bare coupling alone because of the renormalizability of fermion propagator SDE, the bare and renormalized forms of $\alpha\xi$ being identical.
\item for diagonal elements of $\Omega^\Delta$, a trivial solution exists with $\Omega^\Delta=\nu \Theta=-\Omega^\xi$. However, in this case there is no correction to the free-particle propagator.
\end{itemize}
More generally, we define 
\begin{equation}
\Omega^\Delta=
\begin{pmatrix}
\Omega^\Delta_{11} & -\dfrac{m_B}{p^2}\Omega^\Delta_{12} \\ 
-m_B\Omega^\Delta_{21} & \Omega^\Delta_{22}
\end{pmatrix} \label{eq:Omega_Delta_matrix}
\end{equation}
such that $\Omega^\Delta_{ij}$ correspond to dimensionless transforms. In addition, from Eq.~\eqref{eq:k_exponential}, one can easily verify ${\partial_\nu \mathcal{K}_\xi+\Phi\mathcal{K}_\xi=0}$. The similarity of Eq.~\eqref{eq:D_xi_Omega_Delta} to this differential equation for $\mathcal{K}_\xi$ indicates the following expansions for $\Omega^\Delta_{ij}$,
\begin{subequations}
\label{eq:Omegaij_series}
\begin{align}
& \Omega_{ij}^\Delta \,z^\beta=\sum_{m=0}^{+\infty}\dfrac{(-\nu)^m}{m!}\left(\dfrac{4\pi\mu^2}{p^2} \right)^{m\epsilon}\omega_{ij}(\beta,m)\,z^{\beta+m\epsilon},\label{eq:Omega12_series}\\
&\mathrm{for}~(ij)\neq (12),~\mathrm{and}\nonumber\\
& \Omega_{12}^\Delta\, z^\beta=\sum_{m=0}^{+\infty}\dfrac{(-\nu)^m}{m!}\left(\dfrac{4\pi\mu^2}{p^2} \right)^{m\epsilon}\omega_{12}(\beta,m)\,z^{\beta+m\epsilon+1}.\label{eq:Omegaij/12_series}
\end{align}
\end{subequations}
where the expansion coefficients $\omega_{ij}(\beta,m)$ are allowed to implicitly depend on $\epsilon$. The \lq 12' component of $\Omega^\Delta$ is expanded differently from other components to ensure that $\Omega^\Delta$ given by Eq.~\eqref{eq:Omega_Delta_matrix} translates into operations solely on the spectral variables.

With Eq.~\eqref{eq:Omegaij_series}, the left-hand side of Eq.~\eqref{eq:D_xi_Omega_Delta} can be calculated according to Eq.~\eqref{eq:zbeta_Omega_Delta}. Then recurrence relations for $\omega_{ij}(\beta,m)$ are obtained by the comparison of $\mathcal{O}(\nu^m)$ terms in Eq.~\eqref{eq:D_xi_Omega_Delta}. As a result, we have
\begin{widetext}
\begin{subequations}
\label{eq:rec_omega_ij}
\begin{align}
& \quad -\omega_{11}(\beta,m+1)+\Gamma(\epsilon)\dfrac{\Gamma(1-\epsilon)}{\Gamma(1+\epsilon)}\Bigg\{\dfrac{\Gamma(2+\beta)\Gamma(\beta+\epsilon)}{\Gamma(2+\beta-\epsilon)\Gamma(\beta)}\omega_{11}(\beta+\epsilon,m) -\omega_{11}(\beta,m)\dfrac{\Gamma(2+\beta+m\epsilon)\Gamma(\beta+(m+1)\epsilon)}{\Gamma(2+\beta+(m-1)\epsilon)\Gamma(\beta+m\epsilon)} \Bigg\}\nonumber\\[1.0mm]
& =\begin{cases}
\Gamma(1-\epsilon)\dfrac{\Gamma(\beta+\epsilon-1)\Gamma(\beta+1)}{\Gamma(1+\beta-\epsilon)\Gamma(\beta)} & \mathrm{for}~m=0,\\[2.5mm]
-\dfrac{\Gamma(\epsilon)[\Gamma(1-\epsilon)]^2}{\Gamma(1+\epsilon)}\left(\dfrac{\beta+1}{\beta+2\epsilon}-\dfrac{1+\beta+\epsilon}{\beta+1} \right)\dfrac{\Gamma(\beta+2\epsilon)\Gamma(\beta+\epsilon+1)}{\Gamma(\beta-\epsilon+2)\Gamma(\beta)} & \mathrm{for}~m=1,\\[2.5mm]
0 & \mathrm{for}~m\geq 2.
\end{cases}\label{eq:rec_omega_11}
\end{align}
\begin{align}
&\quad  -\omega_{12}(\beta,m+1)+\Gamma(\epsilon)\dfrac{\Gamma(1-\epsilon)}{\Gamma(1+\epsilon)}\Bigg\{\dfrac{\Gamma(2+\beta)\Gamma(\beta+\epsilon)}{\Gamma(2+\beta-\epsilon)\Gamma(\beta)}\omega_{12}(\beta+\epsilon,m)-\omega_{12}(\beta,m)\dfrac{\Gamma(2+\beta+m\epsilon)\Gamma(\beta+1+(m+1)\epsilon)}{\Gamma(2+\beta+(m-1)\epsilon)\Gamma(\beta+1+m\epsilon)}\Bigg\}\nonumber\\[1.0mm]
& =\begin{cases}
\Gamma(1-\epsilon)\dfrac{\Gamma(2+\beta)\Gamma(\beta+\epsilon)}{\Gamma(2+\beta-\epsilon)\Gamma(\beta+1)} & \mathrm{for}~m=0,\\[2.5mm]
 0 & \mathrm{for}~m\geq 1.
\end{cases}\label{eq:rec_omega_12}
\end{align}
\begin{align}
&\quad  -\omega_{21}(\beta,m+1)+\Gamma(\epsilon)\dfrac{\Gamma(1-\epsilon)}{\Gamma(1+\epsilon)}\Bigg\{\dfrac{\Gamma(1+\beta)\Gamma(\beta+\epsilon)}{\Gamma(1+\beta-\epsilon)\Gamma(\beta)}\omega_{21}(\beta+\epsilon,m)-\omega_{21}(\beta,m)\dfrac{\Gamma(2+\beta+m\epsilon)\Gamma(\beta+(m+1)\epsilon)}{\Gamma(2+\beta+(m-1)\epsilon)\Gamma(\beta+m\epsilon)}\Bigg\}\nonumber\\[1.0mm]
&=\begin{cases}
\Gamma(1-\epsilon)\dfrac{\Gamma(1+\beta)\Gamma(\beta+\epsilon)}{\Gamma(2+\beta-\epsilon)\Gamma(\beta)} & \mathrm{for}~m=0,\\[2.5mm]
0 & \mathrm{for}~m\geq 1.
\end{cases}\label{eq:rec_omega_21}
\end{align}
\begin{align}
& \quad -\omega_{22}(\beta,m+1)+\Gamma(\epsilon)\dfrac{\Gamma(1-\epsilon)}{\Gamma(1+\epsilon)}\Bigg\{\dfrac{\Gamma(1+\beta)\Gamma(\beta+\epsilon)}{\Gamma(1+\beta-\epsilon)\Gamma(\beta)}\omega_{22}(\beta+\epsilon,m) -\omega_{22}(\beta,m)\dfrac{\Gamma(1+\beta+m\epsilon)\Gamma(\beta+(m+1)\epsilon)}{\Gamma(1+\beta+(m-1)\epsilon)\Gamma(\beta+m\epsilon)}\Bigg\}\nonumber\\[1.0mm]
& =\begin{cases}
\Gamma(1-\epsilon)\dfrac{\Gamma(\beta+\epsilon)\Gamma(\beta+1)}{\Gamma(2+\beta-\epsilon)\Gamma(\beta)} & \mathrm{for}~m=0,\\[2.5mm]
-\dfrac{\Gamma(\epsilon)[\Gamma(1-\epsilon)]^2}{\Gamma(1+\epsilon)}\left(\dfrac{1}{\beta+1}-\dfrac{1}{1+\beta-\epsilon} \right)\dfrac{\Gamma(\beta+2\epsilon)\Gamma(\beta+\epsilon+1)}{\Gamma(\beta-\epsilon+1)\Gamma(\beta)} & \mathrm{for}~m=1,\\[2.5mm]
0 & \mathrm{for}~m\geq 2.
\end{cases}\label{eq:rec_omega_22}
\end{align}
\end{subequations}
\end{widetext}
These recurrence relations specify how gauge covariance is satisfied when distributions $\Omega^\Delta_{ij}$ are expanded as Taylor series in $\nu=\alpha\xi/4\pi$ written in Eq.~\eqref{eq:Omegaij_series}. On one hand, when the $\Omega^\Delta_{ij}$ are only known in the Landau gauge, Eq.~\eqref{eq:rec_omega_ij} can be used to calculate  $\Omega^\Delta_{ij}$  in any other covariant gauge. On the other hand, when an ans\"atz for $S_F(k)\Gamma^\mu(k,p)S_F(p)$ is known, the operations of $\Omega^\Delta_{ij}$ on $z^\beta$ can be calculated. Eq.~\eqref{eq:rec_omega_ij} then works to verify if this ans\"atz ensures that solutions to fermion propagator SDE are consistent with LKFT.
\subsection{Example: The Gauge Technique in the quenched approximation in 4D}
In the quenched approximation with the Gauge Technique ans\"{a}tz for $S_F(k)\Gamma^\mu(k,p)S_F(p)$ \cite{Delbourgo:1977jc}, we deduce the $\Omega_{ij}$  to be
\begin{align}
\Omega_{11}(s,s';\xi)& =-\dfrac{3\alpha}{4\pi}\bigg\{\left(\dfrac{1}{\epsilon}-\gamma_E+\ln 4\pi+\dfrac{4}{3}+\ln\dfrac{\mu^2}{s}\right)\nonumber\\
& \quad\times\delta(s-s')-\dfrac{s'}{s^2}\theta(s-s') \bigg\}-\dfrac{\alpha\xi}{4\pi}\dfrac{1}{s}\theta(s-s'),\nonumber\\[1mm]
\Omega_{12}(s,s';\xi)& =-\dfrac{m_B}{s}\delta(s-s'),\nonumber\\[1mm]
\Omega_{21}(s,s';\xi)& =-m_B \delta(s-s'),\nonumber\\[1mm]
\Omega_{22}(s,s';\xi)&=-\dfrac{3\alpha}{4\pi}\bigg\{\left(\dfrac{1}{\epsilon}-\gamma_E+\ln 4\pi+\dfrac{4}{3}+\ln\dfrac{\mu^2}{s}\right)\nonumber\\
& \quad\times\delta(s-s')-\dfrac{1}{s}\theta(s-s') \bigg\}-\dfrac{\alpha\xi}{4\pi}\dfrac{s'}{s^2}\theta(s-s'),\label{eq:Omega_GT}
\end{align}
where $m_B$ is the bare mass and $d=4-2\epsilon$. Equivalently written as operators on $z$, $\Omega_{ij}$ become
\begin{align}
\Omega_{11}(\xi)& =-\dfrac{3\alpha}{4\pi}\left[\tilde{C}+\ln(z) -z^{-1}I\right]-\dfrac{\alpha\xi}{4\pi}Iz^{-1},\nonumber\\[1mm]
\Omega_{12}& =-\dfrac{m_B}{p^2}z,\nonumber\\
\Omega_{21}& =-m_B,\nonumber\\[1mm]
\Omega_{22}(\xi)& =-\dfrac{3\alpha}{4\pi}\left[\tilde{C}+\ln(z) -Iz^{-1}\right]-\dfrac{\alpha\xi}{4\pi}z^{-1}I,\label{eq:Omega_GT_z}
\end{align}
where $\tilde{C}=1/\epsilon-\gamma_E+\ln(4\pi\mu^2/p^2)+4/3$.

Meanwhile, since in 4D the LKFT for the fermion propagator reduces to Eq.~\eqref{eq:kn_small_epsilon}, we have, 
\begin{subequations}
\begin{align}
z^\beta \mathcal{K}_1(\xi)& =\left(\dfrac{\mu^2}{p^2}\right)^{-\nu}\exp\bigg\{-\nu\left[ \dfrac{1}{\epsilon}+\gamma_E+\ln(4\pi)\right]\bigg\}\nonumber\\
& \quad\times\dfrac{\Gamma(\beta-\nu)\Gamma(2+\beta-\nu)}{\Gamma(\beta)\Gamma(2+\beta)}z^{\beta-\nu},\\[1.5mm]
z^\beta \mathcal{K}_2(\xi)& =\left(\dfrac{\mu^2}{p^2}\right)^{-\nu}\exp\bigg\{-\nu\left[ \dfrac{1}{\epsilon}+\gamma_E+\ln(4\pi)\right]\bigg\}\nonumber\\
& \quad\times\dfrac{\Gamma(\beta-\nu)\Gamma(1+\beta-\nu)}{\Gamma(\beta)\Gamma(1+\beta)}z^{\beta-\nu}.
\end{align}
\end{subequations}
For the consistency requirement, it is more convenient to write Eq.~\eqref{eq:consistency_fermion_prop_SDE_LKFT} as 
\begin{align}
&\quad 
\begin{pmatrix}
\Omega_{11}(\xi) & \Omega_{12}(\xi) \\ 
\Omega_{21}(\xi) & \Omega_{22}(\xi)
\end{pmatrix}\nonumber\\
& =
\begin{pmatrix}
\mathcal{K}_1 (\xi)\Omega_{11}(0)\mathcal{K}_1(-\xi) & \mathcal{K}_1 (\xi)\Omega_{12}(0)\mathcal{K}_2(-\xi) \\ 
\mathcal{K}_2 (\xi)\Omega_{21}(0)\mathcal{K}_1(-\xi) & \mathcal{K}_2(\xi)\Omega_{22}(0)\mathcal{K}_2(-\xi)
\end{pmatrix} .\label{eq:consistency_fermion_prop_SDE_LKFT_inverse}
\end{align}
With the assistance of the following four identities for fractional calculus, 
\begin{subequations}
\begin{align}
I^\alpha z^\beta &=\dfrac{\Gamma(\beta+1)}{\Gamma(\alpha+\beta+1)}z^{\alpha+\beta},\\
D^\alpha z^\beta &=\dfrac{\Gamma(\beta+1)}{\Gamma(-\alpha+\beta+1)}z^{-\alpha+\beta},\\
I^\alpha z^\beta \ln(z)& =\dfrac{\Gamma(\beta+1)}{\Gamma(\alpha+\beta+1)}\big\{\psi(\beta+1)-\nonumber\\
& \quad\psi(\alpha+\beta+1) +\ln(z) \big\}z^{\alpha+\beta},\\
D^\alpha z^\beta \ln(z)& =\dfrac{\Gamma(\beta+1)}{\Gamma(-\alpha+\beta+1)}\big\{\psi(\beta+1)\nonumber\\
& \quad-\psi(-\alpha+\beta+1) +\ln(z)\big\}z^{-\alpha+\beta},
\end{align}
\end{subequations}
where $\psi(\beta)$ is the digamma function, one then obtains
\begin{subequations}
\label{eq:K_Omega0_Kinv_zbeta}
\begin{align}
& \quad z^\beta \mathcal{K}_1(\xi)\Omega_{11}(0)\mathcal{K}_1(-\xi)\nonumber\\
& =-\dfrac{3\alpha}{4\pi}\bigg\{\tilde{C}-\dfrac{1}{\beta-\nu+1}+\psi(\beta)-\psi(\beta-\nu)+\nonumber\\
& \hspace{2cm}\psi(\beta+2) -\psi(\beta+2-\nu)+\ln z \bigg\}z^\beta,
\end{align}
\begin{align}
& z^\beta \mathcal{K}_1(\xi)\Omega_{12}(0)\mathcal{K}_2(-\xi)=-\dfrac{m_B}{p^2}\dfrac{\beta}{\beta-\nu}z^{\beta+1},\\[3mm]
& z^\beta \mathcal{K}_2(\xi)\Omega_{21}(0)\mathcal{K}_1(-\xi)=-m_B\dfrac{\beta+1}{\beta+1-\nu}z^\beta,
\end{align}
\begin{align}
& z^\beta \mathcal{K}_2(\xi)\Omega_{22}(0)\mathcal{K}_2(-\xi)\nonumber\\
& =-\dfrac{3\alpha}{4\pi}\bigg\{\tilde{C}-\dfrac{1}{\beta-\nu}+\psi(\beta)-\psi(\beta-\nu)+\nonumber\\
& \hspace{2cm} \psi(\beta+1)-\psi(\beta+1-\nu)+\ln z \bigg\}z^\beta.
\end{align}
\end{subequations}
 While from Eq.~\eqref{eq:Omega_GT_z}, we have
\begin{subequations}
\label{eq:Omega_xi_zbeta}
\begin{align}
& z^\beta \Omega_{11}(\xi)=\bigg\{-\dfrac{3\alpha}{4\pi}\left[\tilde{C}-\dfrac{1}{\beta+1}+\ln z\right]-\dfrac{\nu}{\beta}\bigg\}z^{\beta},\\
& z^\beta\Omega_{12}(\xi)=-\dfrac{m_B}{p^2}z^{\beta+1},\\
& z^\beta\Omega_{21}(\xi)=-m_B z^\beta,\\
& z^\beta\Omega_{22}(\xi)=\bigg\{-\dfrac{3\alpha}{4\pi}\left[\tilde{C}-\dfrac{1}{\beta}+\ln z\right]-\dfrac{\nu}{\beta+1}\bigg\}z^{\beta}
\end{align}
\end{subequations}
Observe that the digamma functions only occur in  Eq.~\eqref{eq:K_Omega0_Kinv_zbeta},  not in Eq.~\eqref{eq:Omega_xi_zbeta}. Additionally, the dependence on $\nu$ is only linear in Eq.~\eqref{eq:Omega_xi_zbeta}, but not in Eq.~\eqref{eq:K_Omega0_Kinv_zbeta}. Therefore the consistency requirement given by Eq.~\eqref{eq:consistency_fermion_prop_SDE_LKFT_inverse} is not satisfied by the Gauge Technique in 4D. The same conclusion has been realized by Delbourgo, Keck and Parker \cite{Delbourgo:1980vc} in a completely different approach.
\section{Summary\label{ss:summary}}
In this paper we have formulated the fermion propagator SDE in terms of propagator spectral functions. With the fermion-photon vertex structure $S_F(k)\Gamma^\mu(k,p)S_F(p)$ being linear in the $\rho_j(s;\xi)$ as implied by the equality of renormalization factors $Z_1=Z_2$, we have derived the necessary and sufficient condition for the solutions of the fermion propagator SDE to be consistent with LKFT in covariant gauges. With known contributions to the fermion propagator SDE calculated, this reduces the consistency requirement to that for the contribution to $\Omega$ in Eq.~\eqref{eq:SDE_fermion_rho_itg} from the Landau gauge photon propagator. 
Next, an expansion of the operator $\Omega^\Delta$ (defined in Eq.\eqref{eq:Omega_decomposition}), similar to that
of $\mathcal{K}_j$ in Eq.~\eqref{eq:kn_zbeta}, has been postulated in Eq.~\eqref{eq:Omegaij_series}. The consistency
requirements can then be converted into the form of recurrence relations of this expansion,
shown in Eq.~\eqref{eq:rec_omega_ij}.
The requirement on $S_F(k)\Gamma^\mu(k,p)S_F(p)$ to ensure the gauge invariance of $\Pi(q^2)$ was also derived. 

We observe that the Gauge Technique \cite{PhysRev.130.1287,PhysRev.135.B1398,PhysRev.135.B1428,Delbourgo:1977jc} does not ensure gauge covariance for the fermion propagator in QED. In fact, when fermions are massive, dimension-odd operators are required in $S_F(k)\Gamma^\mu(k,p)S_F(p)$ to ensure gauge covariance. Our formalism for the SDEs using a spectral representation allows propagators to be solved in Minkowski space. Furthermore, our consistency requirements can be used as criteria for truncating the SDEs for QED propagators. 

Importantly, our calculations have been performed in arbitary dimensions.  Keeping $\epsilon = 2- d/2$ explicit to the end turns out to give concise and meaningful results in the case of the $\sigma_{j}^{\xi}$ in Eq.~\eqref{eq:def_sigma_overline_Delta}, 
the fermion Schwinger-Dyson equation, as well as the LKFT for the fermion propagator. Results are concise in the sense that one hypergeometric function describes the $p^2$ dependence for each Dirac component of every loop integral. Meaningful in the sense that the results apply to any number of spacetime dimensions as long as hypergeometric functions converge. Based on these two merits, one might suspect that dimensional regularization evaluated by keeping $\epsilon$ explicit to the last step is intrinsic to QED itself.

This work marks a path towards ensuring consistent truncations of the Schwinger-Dyson equations for the fermion and boson propagators yield gauge covariant fermion mass functions like that in Fig.~1: an essential requirement for validating any truncation scheme used.
\begin{acknowledgments}
This material is based upon work supported by the U.S. Department of Energy, Office of Science, Office of Nuclear Physics under contract DE-AC05-06OR23177 that funds Jefferson Lab research. The authors would like to thank Professor Keith Ellis and other members of the Institute for Particle Physics Phenomenology (IPPP) of Durham University for kind hospitality during their visit when this article was finalized.
\end{acknowledgments}
\appendix
\section{Simplification of $\sigma_j^\xi(p^2;\xi)$\label{ss:simplify_sigma_xi}}
To simplify Eq.~\eqref{eq:sigma_overline_delta_loop}, we will need contiguous relations for hypergeometric functions from Ref.~\cite{abramowitz1964handbook} and the following identity
\begin{align}
&\quad (A\slashed{p}+BW)(\slashed{p}+W)=s(Az+B)+(A+B)\slashed{p}W\nonumber\\
&=s(z-1)\left[\left(A+\dfrac{A+B}{z-1} \right)+\dfrac{A+B}{s(z-1)}\slashed{p}W\right].
\end{align}
Equations referred to by Eq.~(15.2.XX) are identities in Ref.~\cite{abramowitz1964handbook}. With ${ a=3,~b=1+\epsilon,~c=4-\epsilon}$, Eq.~(15.2.19) becomes
\begin{align}
&\quad \dfrac{2(3-2\epsilon)}{(3-\epsilon)(2-\epsilon)(1-\epsilon)}~_2F_1(\epsilon,3;4-\epsilon;z)\nonumber\\
& =\dfrac{2}{(3-\epsilon)(2-\epsilon)}~_2F_1(2,1+\epsilon;4-\epsilon;z)+\nonumber\\
& \quad \dfrac{2(1-z)}{(3-\epsilon)(1-\epsilon)}~_2F_1(3,1+\epsilon;4-\epsilon;z).\label{eq:15.2.19.1}
\end{align}
Explicitly then,
\begin{align}
A & =\dfrac{-\epsilon}{(3-\epsilon)(2-\epsilon)}~_2F_1(1+\epsilon,2;4-\epsilon;z)+ \nonumber\\
& \quad \dfrac{3\epsilon-4}{(2-\epsilon)(1-\epsilon)}~_2F_1(\epsilon,2;3-\epsilon;z)+ \nonumber\\
& \quad \dfrac{2(3-2\epsilon)}{(3-\epsilon)(2-\epsilon)(1-\epsilon)}~_2F_1(\epsilon,3;4-\epsilon;z) \nonumber\\
& =\dfrac{2(1-z)}{(3-\epsilon)(1-\epsilon)}~_2F_1(1+\epsilon,3;4-\epsilon;z)+\label{eq:15.2.19.1.a}\\
& \quad \dfrac{1}{3-\epsilon}~_2F_1(1+\epsilon,2;4-\epsilon;z)+ \nonumber\\
& \quad \dfrac{3\epsilon-4}{(2-\epsilon)(1-\epsilon)}~_2F_1(\epsilon,2;3-\epsilon;z),\nonumber
\end{align}
where Eq.~\eqref{eq:15.2.19.1} is used to derive Eq.~\eqref{eq:15.2.19.1.a}.
From Eq.~(15.2.17) with $a=1,~b=\epsilon,~c=3-\epsilon$ we have
\begin{align}
B& =\dfrac{1}{1-\epsilon}~_2F_1(\epsilon,1;2-\epsilon;z)\nonumber\\
& =\dfrac{1}{2-\epsilon}~_2F_1(\epsilon,1;3-\epsilon;z)+\nonumber\\
& \quad \dfrac{1}{(2-\epsilon)(1-\epsilon)}~_2F_1(\epsilon,2;3-\epsilon;z).
\end{align}
Next, with ${a=\epsilon,~b=2,~c=3-\epsilon}$, Eq.~(15.2.15) becomes
\begin{align}
& \quad\dfrac{1}{2-\epsilon}~_2F_1(\epsilon,1;3-\epsilon;z) \nonumber\\
&  =\dfrac{1-2\epsilon}{(2-\epsilon)(1-\epsilon)}~_2F_1(2,\epsilon;3-\epsilon;z)+\nonumber\\
& \quad \dfrac{\epsilon(1-z)}{(2-\epsilon)(1-\epsilon)}~_2F_1(2,\epsilon+1;3-\epsilon;z).\label{eq:15.2.15.1}
\end{align}
With ${a=\epsilon,~b=3,~c=4-\epsilon}$, Eq.~(15.2.17) becomes
\begin{align}
& \quad \dfrac{-1}{1-\epsilon}~_2F_1(\epsilon,2;3-\epsilon;z) \nonumber\\
&=\dfrac{-(3-2\epsilon)}{(3-\epsilon)(1-\epsilon)}~_2F_1(\epsilon,2;4-\epsilon;z)+\nonumber\\
& \quad \dfrac{-\epsilon}{(3-\epsilon)(1-\epsilon)}~_2F_1(\epsilon+1,2;4-\epsilon;z). \label{eq:15.2.17.1}
\end{align}
With ${a=2,~b=1+\epsilon,~c=4-\epsilon}$, Eq.~(15.2.15) becomes
\begin{align}
& \quad -(3-2\epsilon)~_2F_1(\epsilon,2;4-\epsilon;z)+\nonumber\\
& \quad (1-2\epsilon)~_2F_1(\epsilon+1,2;4-\epsilon;z)\nonumber\\
& =-2(1-z)~_2F_1(\epsilon+1,3;4-\epsilon;z).\label{eq:15.2.15.2}
\end{align}
Therefore
\begin{subequations}
\begin{align}
& \quad A+B \nonumber\\
& =\dfrac{2(1-z)}{(3-\epsilon)(1-\epsilon)}~_2F_1(1+\epsilon,3;4-\epsilon;z)+ \label{eq:15.2.15.1.a}\\
& \quad \dfrac{\epsilon(1-z)}{(2-\epsilon)(1-\epsilon)}~_2F_1(\epsilon+1,2;3-\epsilon;z)+ \nonumber\\
& \quad\dfrac{-1}{1-\epsilon}~_2F_1(\epsilon,2;3-\epsilon;z)+ \nonumber\\
& \quad \dfrac{1}{3-\epsilon}~_2F_1(1+\epsilon,2;4-\epsilon;z) \nonumber\\
& =\dfrac{2(1-\epsilon)}{(3-\epsilon)(1-\epsilon)}~_2F_1(1+\epsilon,3;4-\epsilon;z)+ \label{eq:15.2.17.1.a}\\
&\quad \dfrac{\epsilon(1-z)}{(2-\epsilon)(1-\epsilon)}~_2F_1(\epsilon+1,2;3-\epsilon;z)+ \nonumber\\
& \quad\dfrac{-(3-2\epsilon)}{(3-\epsilon)(1-\epsilon)}~_2F_1(\epsilon,2;4-\epsilon;z)+ \nonumber\\
& \quad \dfrac{1-2\epsilon}{(3-\epsilon)(1-\epsilon)}~_2F_1(\epsilon+1,2;4-\epsilon;z) \nonumber\\
& =\dfrac{\epsilon(1-z)}{(2-\epsilon)(1-\epsilon)}~_2F_1(\epsilon+1,2;3-\epsilon;z), \label{eq:15.2.15.2.a}
\end{align}
\end{subequations}
where Eqs.~(\ref{eq:15.2.15.1},~\ref{eq:15.2.17.1},~\ref{eq:15.2.15.2}) are used to derive Eqs.~(\ref{eq:15.2.15.1.a},~\ref{eq:15.2.17.1.a},~\ref{eq:15.2.15.2.a}), respectively. In addition, with ${a=\epsilon,~b=2,~c=4-\epsilon}$, Eq.~(15.2.14) becomes
\begin{align}
& \quad \epsilon ~_2F_1(\epsilon+1,2;4-\epsilon;z) \nonumber\\
& =2~_2F_1(\epsilon,3;4-\epsilon;z)- (2-\epsilon)~_2F_1(\epsilon,2;4-\epsilon;z).\label{eq:15.2.14.1}
\end{align}
With ${a=\epsilon,~b=2,~c=3-\epsilon}$, Eq.~(15.2.14) becomes
\begin{align}
&  \quad\epsilon ~_2F_1(\epsilon+1,2;3-\epsilon;z)\nonumber\\
& =2~_2F_1(\epsilon,3;3-\epsilon;z)- (2-\epsilon)~_2F_1(\epsilon,2;4-\epsilon;z).\label{eq:15.2.14.2}
\end{align}
With ${a=\epsilon,~b=2,~c=4-\epsilon}$, Eq.~(15.2.24) becomes
\begin{align}
& \quad (1-\epsilon)~_2F_1(\epsilon,2;4-\epsilon;z)+ 2~_2F_1(\epsilon,3;4-\epsilon;z) \nonumber\\
& =(3-\epsilon)~_2F_1(\epsilon,2;3-\epsilon;z).\label{eq:15.2.24.1}
\end{align}
With ${a=\epsilon,~b=2,~c=3-\epsilon}$, Eq.~(15.2.24) becomes
\begin{align}
& \quad -\epsilon ~_2F_1(\epsilon,2;3-\epsilon;z)+2~_2F_1(\epsilon,2;3-\epsilon;z)\nonumber\\
&  =(2-\epsilon)~_2F_1(\epsilon,2;2-\epsilon;z).\label{eq:15.2.24.2}
\end{align}
Then
\begin{widetext}
\begin{subequations}
\begin{align}
A + \frac{A+B}{(z-1)}& =\dfrac{-\epsilon}{(3-\epsilon)(2-\epsilon)}~_2F_1(1+\epsilon,2;4-\epsilon;z)+\dfrac{3\epsilon-4}{(2-\epsilon)(1-\epsilon)}~_2F_1(\epsilon,2;3-\epsilon;z)+ \nonumber\\
& \quad \dfrac{2(3-2\epsilon)}{(3-\epsilon)(2-\epsilon)(1-\epsilon)}~_2F_1(\epsilon,3;4-\epsilon;z)+  \dfrac{-\epsilon}{(2-\epsilon)(1-\epsilon)}~_2F_1(\epsilon+1,2;3-\epsilon;z) \nonumber\\
& =\dfrac{1}{(3-\epsilon)(2-\epsilon)}\bigg\{\dfrac{2(2-\epsilon)}{1-\epsilon}~_2F_1(\epsilon,3;4-\epsilon;z)+(2-\epsilon)~_2F_1(\epsilon,2;4-\epsilon;z)\bigg\}+ \label{eq:15.2.14.1a}\\
& \quad \dfrac{1}{(2-\epsilon)(1-\epsilon)}\big\{-2 ~_2F_1(\epsilon,3;3-\epsilon;z)+ 2(\epsilon-1)~_2F_1(\epsilon,2;3-\epsilon;z)\big\}\nonumber\\
& = \dfrac{2}{(3-\epsilon)(1-\epsilon)}~_2F_1(\epsilon,3;4-\epsilon;z)+ \dfrac{1}{3-\epsilon}~_2F_1(\epsilon,2;4-\epsilon;z)+\label{eq:15.2.14.2a}\\
& \quad\dfrac{-2}{(2-\epsilon)(1-\epsilon)}~_2F_1(\epsilon,3;3-\epsilon;z)+ \dfrac{-2}{2-\epsilon}~_2F_1(\epsilon,2;3-\epsilon;z) \nonumber\\
& =\dfrac{\epsilon}{(1-\epsilon)(2-\epsilon)}~_2F_1(\epsilon,2;3-\epsilon;z)+ \dfrac{-2}{(2-\epsilon)(1-\epsilon)}~_2F_1(\epsilon,3;3-\epsilon;z) \label{eq:15.2.24.1a}\\
& =\dfrac{-1}{1-\epsilon}~_2F_1(\epsilon,2;2-\epsilon;z), \label{eq:15.2.24.2a}
\end{align}
\end{subequations}
where Eqs.~(\ref{eq:15.2.14.1},~\ref{eq:15.2.14.2},~\ref{eq:15.2.24.1},~\ref{eq:15.2.24.2}) have been utilized to derive Eqs.~(\ref{eq:15.2.14.1a},~\ref{eq:15.2.14.2a},~\ref{eq:15.2.24.1a},~\ref{eq:15.2.24.2a}), respectively.
Finally we obtain,
\begin{align}
ie^2\xi\int d\underline{k}\slashed{q}\dfrac{1}{\slashed{k}-W}\dfrac{\slashed{q}}{q^4}\dfrac{1}{\slashed{p}-W}&=\dfrac{-\alpha\xi}{4\pi}\Gamma(\epsilon)\left(\dfrac{4\pi\mu^2}{s} \right)^\epsilon \Bigg\{\dfrac{-1}{1-\epsilon}~_2F_1(\epsilon,2;2-\epsilon;z)+\nonumber\\
& \quad\hspace{4cm} \dfrac{-\epsilon}{(2-\epsilon)(1-\epsilon)}~_2F_1(\epsilon+1,2;3-\epsilon;z)\dfrac{\slashed{p}}{W} \Bigg\}.
\end{align}
\section{Operations on $z^\beta$ from terms in Eq.~\eqref{eq:D_xi_Omega_Delta}\label{ss:recurrence_z_beta}}
For commutators on the right-hand side of Eq.~\eqref{eq:D_xi_Omega_Delta}, explicit calculation shows that
\begin{subequations}\label{eq:zbeta_phi_theta}
\begin{align}
& z^\beta \phi_3\theta_1=\dfrac{\Gamma(\epsilon)[\Gamma(1-\epsilon)]^2}{\Gamma(1+\epsilon)}\left(\dfrac{4\pi\mu^2}{p^2} \right)^{2\epsilon} \dfrac{\beta+1}{\beta+2\epsilon} \dfrac{\Gamma(\beta+2\epsilon)\Gamma(\beta+\epsilon+1)}{\Gamma(\beta-\epsilon+2)\Gamma(\beta)}z^{\beta+2\epsilon},\\[1mm]
& z^\beta \theta_1\phi_3=\dfrac{\Gamma(\epsilon)[\Gamma(1-\epsilon)]^2}{\Gamma(1+\epsilon)}\left(\dfrac{4\pi\mu^2}{p^2} \right)^{2\epsilon} \dfrac{1+\beta+\epsilon}{\beta+1} \dfrac{\Gamma(\beta+2\epsilon)\Gamma(\beta+\epsilon+1)}{\Gamma(\beta-\epsilon+2)\Gamma(\beta)}z^{\beta+2\epsilon},\\[1mm]
& z^\beta \phi_2\theta_2=\dfrac{\Gamma(\epsilon)[\Gamma(1-\epsilon)]^2}{\Gamma(1+\epsilon)}\left(\dfrac{4\pi\mu^2}{p^2} \right)^{2\epsilon} \dfrac{1}{\beta+1} \dfrac{\Gamma(\beta+2\epsilon)\Gamma(\beta+\epsilon+1)}{\Gamma(\beta-\epsilon+1)\Gamma(\beta)}z^{\beta+2\epsilon},\\[1mm]
& z^\beta \theta_2\phi_2=\dfrac{\Gamma(\epsilon)[\Gamma(1-\epsilon)]^2}{\Gamma(1+\epsilon)}\left(\dfrac{4\pi\mu^2}{p^2} \right)^{2\epsilon} \dfrac{1}{1+\beta-\epsilon} \dfrac{\Gamma(\beta+2\epsilon)\Gamma(\beta+\epsilon+1)}{\Gamma(\beta-\epsilon+1)\Gamma(\beta)}z^{\beta+2\epsilon},
\end{align}
\end{subequations}
and
\begin{subequations}\label{eq:zbeta_phi}
\begin{align}
z^\beta(\phi_3z-z\phi_3)& =\Gamma(\epsilon)\left(\dfrac{4\pi\mu^2}{p^2} \right)^\epsilon \dfrac{\Gamma(1-\epsilon)}{\Gamma(1+\epsilon)}\left[\dfrac{\Gamma(2+\beta)\Gamma(\beta+\epsilon)}{\Gamma(2+\beta-\epsilon)\Gamma(\beta)}-\dfrac{\Gamma(2+\beta)\Gamma(\beta+1+\epsilon)}{\Gamma(2+\beta-\epsilon)\Gamma(\beta+1)} \right]z^{\beta+\epsilon+1}\nonumber\\[1mm]
& =-\Gamma(1-\epsilon)\left(\dfrac{4\pi\mu^2}{p^2} \right)^\epsilon\dfrac{\Gamma(2+\beta)\Gamma(\beta+\epsilon)}{\Gamma(2+\beta-\epsilon)\Gamma(\beta+1)}z^{\beta+\epsilon+1}\\[1mm]
z^\beta(\phi_2-\phi_3)& =\Gamma(\epsilon)\left(\dfrac{4\pi\mu^2}{p^2} \right)^\epsilon \dfrac{\Gamma(1-\epsilon)}{\Gamma(1+\epsilon)}\left[\dfrac{\Gamma(1+\beta)\Gamma(\beta+\epsilon)}{\Gamma(1+\beta-\epsilon)\Gamma(\beta)}-\dfrac{\Gamma(2+\beta)\Gamma(\beta+\epsilon)}{\Gamma(2+\beta-\epsilon)\Gamma(\beta)} \right]z^{\beta+\epsilon}\nonumber\\[1mm]
& =-\Gamma(1-\epsilon)\left(\dfrac{4\pi\mu^2}{p^2} \right)^\epsilon\dfrac{\Gamma(1+\beta)\Gamma(\beta+\epsilon)}{\Gamma(2+\beta-\epsilon)\Gamma(\beta)}z^{\beta+\epsilon}.
\end{align}
\end{subequations}
Up until now all terms on the right-hand side of Eq.~\eqref{eq:D_xi_Omega_Delta} are explicit. For the left-hand side, we have
\begin{subequations}\label{eq:zbeta_Omega_Delta}
\begin{align}
z^\beta\partial_\nu \Omega_{ij}^\Delta&=\sum_{m=0}^{+\infty}\dfrac{(-\nu)^m}{m!}\left(\dfrac{4\pi\mu^2}{p^2}\right)^{(m+1)\epsilon}[-\omega_{ij}(\beta,m+1)]z^{\beta+(m+1)\epsilon}\quad (i,j)\neq (1,2),\\[1mm]
z^\beta\partial_\nu \Omega_{12}^\Delta&=\sum_{m=0}^{+\infty}\dfrac{(-\nu)^m}{m!}\left(\dfrac{4\pi\mu^2}{p^2} \right)^{(m+1)\epsilon}\left[-\omega_{12}(\beta,m+1)\right]z^{1+\beta+(m+1)\epsilon}\\[1mm]
z^\beta(\phi_3\Omega_{11}^\Delta-\Omega_{11}^\Delta\phi_3)& =\sum_{m=0}^{+\infty}\dfrac{(-\nu)^m}{m!}\left(\dfrac{4\pi\mu^2}{p^2} \right)^{(m+1)\epsilon}\Gamma(\epsilon)\dfrac{\Gamma(1-\epsilon)}{\Gamma(1+\epsilon)}\Bigg\{\dfrac{\Gamma(2+\beta)\Gamma(\beta+\epsilon)}{\Gamma(2+\beta-\epsilon)\Gamma(\beta)}\omega_{11}(\beta+\epsilon,m)\nonumber\\[1mm]
& \quad -\omega_{11}(\beta,m)\dfrac{\Gamma(2+\beta+m\epsilon)\Gamma(\beta+(m+1)\epsilon)}{\Gamma(2+\beta+(m-1)\epsilon)\Gamma(\beta+m\epsilon)} \Bigg\}z^{\beta+(m+1)\epsilon}\\[1mm]
z^\beta(\phi_3\Omega_{12}^\Delta-\Omega_{12}^\Delta\phi_2)& =\sum_{m=0}^{+\infty}\dfrac{(-\nu)^m}{m!}\left(\dfrac{4\pi\mu^2}{p^2} \right)^{(m+1)\epsilon}\Gamma(\epsilon)\dfrac{\Gamma(1-\epsilon)}{\Gamma(1+\epsilon)}\Bigg\{\dfrac{\Gamma(2+\beta)\Gamma(\beta+\epsilon)}{\Gamma(2+\beta-\epsilon)\Gamma(\beta)}\omega_{12}(\beta+\epsilon,m)\nonumber\\[1mm]
& \quad -\omega_{12}(\beta,m)\dfrac{\Gamma(2+\beta+m\epsilon)\Gamma(\beta+1+(m+1)\epsilon)}{\Gamma(2+\beta+(m-1)\epsilon)\Gamma(\beta+1+m\epsilon)}\Bigg\}z^{1+\beta+(m+1)\epsilon}\\[1mm]
z^\beta(\phi_2\Omega_{21}^\Delta-\Omega_{21}^\Delta\phi_3) &=\sum_{m=0}^{+\infty}\dfrac{(-\nu)^m}{m!}\left(\dfrac{4\pi\mu^2}{p^2} \right)^{(m+1)\epsilon}\Gamma(\epsilon)\dfrac{\Gamma(1-\epsilon)}{\Gamma(1+\epsilon)}\Bigg\{\dfrac{\Gamma(1+\beta)\Gamma(\beta+\epsilon)}{\Gamma(1+\beta-\epsilon)\Gamma(\beta)}\omega_{21}(\beta+\epsilon,m)\nonumber\\[1mm]
& \quad -\omega_{21}(\beta,m)\dfrac{\Gamma(2+\beta+m\epsilon)\Gamma(\beta+(m+1)\epsilon)}{\Gamma(2+\beta+(m-1)\epsilon)\Gamma(\beta+m\epsilon)}\Bigg\}z^{\beta+(m+1)\epsilon}\\[1mm]
z^\beta(\phi_2\Omega_{22}^\Delta-\Omega_{22}^\Delta\phi_2)& =\sum_{m=0}^{+\infty}\dfrac{(-\nu)^m}{m!}\left(\dfrac{4\pi\mu^2}{p^2} \right)^{(m+1)\epsilon}\Gamma(\epsilon)\dfrac{\Gamma(1-\epsilon)}{\Gamma(1+\epsilon)}\Bigg\{\dfrac{\Gamma(1+\beta)\Gamma(\beta+\epsilon)}{\Gamma(1+\beta-\epsilon)\Gamma(\beta)}\omega_{22}(\beta+\epsilon,m)\nonumber\\[1mm]
& \quad -\omega_{22}(\beta,m)\dfrac{\Gamma(1+\beta+m\epsilon)\Gamma(\beta+(m+1)\epsilon)}{\Gamma(1+\beta+(m-1)\epsilon)\Gamma(\beta+m\epsilon)}\Bigg\}z^{\beta+(m+1)\epsilon}.
\end{align}
\end{subequations}
\end{widetext}
\bibliographystyle{utphys}
\bibliography{GC_fSDE_QED_bib}
\end{document}